\documentclass[12pt]{article}
\usepackage{amssymb} 
\usepackage{graphics} 
\usepackage{cite}
\usepackage{epsfig}
\newcommand  {\Bioch}    {{\it Biochemistry\ }}

\newcommand  {\BJ}       {{\it Biophys.~J.\ }}

\newcommand  {\COSB}     {{\it Curr.\ Opin.\ Struct.\ Biol.\ }}

\newcommand  {\EL}       {{\it Europhys.\ Lett.\ }}

\newcommand  {\JACS}     {{\it J.\ Am.\ Chem.\ Soc.\ }}

\newcommand  {\JCP}      {{\it J.\ Chem.\ Phys.\ }}
\newcommand  {\JMB}      {{\it J.\ Mol.\ Biol.\ }}

\newcommand  {\Nat}      {{\it Nature\ }}

\newcommand  {\NSB}      {{\it Nat.\ Struct.\ Biol.\ }}

\newcommand  {\Pro}      {{\it Proteins\ }}

\newcommand  {\ProSci}   {{\it Protein\ Sci.\ }}

\newcommand  {\PNAS}     {{\it Proc.\ Natl.\ Acad.\ Sci.\ USA\ }}

\newcommand  {\PRL}      {{\it Phys.\ Rev.\ Lett.\ }}

\newcommand  {\Sci}      {{\it Science\ }}

\newcommand  {\TBS}      {{\it Trends Biochem. Sci.\ }}

\newcommand{\beq}{\begin{equation}}
\newcommand{\eeq}{\end{equation}}
\newcommand{\beqa}{\begin{eqnarray}}
\newcommand{\eeqa}{\end{eqnarray}}
\newcommand{\bea}{\begin{eqnarray}}
\newcommand{\eea}{\end{eqnarray}}
\newcommand   {\ev}[1]   {\langle #1\rangle}
\newcommand   {\Nhb}     {N_{\mbox{{\scriptsize hb}}}^{\mbox{{\scriptsize nat}}}}
\newcommand   {\Aba}     {A$\beta_{16-22}$}
\newcommand   {\Fs}      {F${}_{\mbox{{\scriptsize s}}}$}               
\newcommand   {\Ca}      {C${}_{\alpha}$}
\newcommand   {\Cb}      {C${}_{\beta}$}
\newcommand   {\Cd}      {C${}_{\delta}$}
\newcommand   {\Cg}      {C${}_{\gamma}$}
\newcommand   {\Cp}      {C${}^{\prime}$}
\newcommand   {\rc}      {r^{\mbox{{\scriptsize c}}}}                           \newcommand   {\Eloc}    {E_{\mbox{{\scriptsize loc}}}}
\newcommand   {\Eev}     {E_{\mbox{{\scriptsize ev}}}}
\newcommand   {\Ehb}     {E_{\mbox{{\scriptsize hb}}}}

\newcommand   {\Ehp}     {E_{\mbox{{\scriptsize hp}}}}
\newcommand   {\Tm}      {T_{\mbox{{\scriptsize m}}}}

\newcommand   {\Tmin}    {T_{\mbox{{\scriptsize min}}}}
\newcommand   {\Tmax}    {T_{\mbox{{\scriptsize max}}}}
\newcommand   {\dE}      {\Delta E}

\newcommand   {\Cv}      {C_{\mbox{{\scriptsize v}}}}
\newcommand   {\kev}     {\kappa_{\mbox{{\scriptsize ev}}}}
\newcommand   {\kloc}    {\kappa_{\mbox{{\scriptsize loc}}}}

\newcommand   {\ehba}    {\epsilon^{(1)}_{\mbox{{\scriptsize hb}}}}
\newcommand   {\ehbb}    {\epsilon^{(2)}_{\mbox{{\scriptsize hb}}}}
\newcommand   {\Mij}     {M_{IJ}}                                               \newcommand   {\shb}     {\sigma_{\mbox{{\scriptsize hb}}}}

\newcommand   {\Rg}      {R_{\mbox{{\scriptsize g}}}}
\newcommand   {\Db}      {\Delta_{\mbox{{\scriptsize b}}}}
\newcommand   {\Xn}      {X_{\mbox{\scriptsize n}}}
\newcommand   {\Xu}      {X_{\mbox{\scriptsize u}}}

\newcommand   {\En}      {E_{\mbox{\scriptsize n}}}
\newcommand   {\Eu}      {E_{\mbox{\scriptsize u}}}

\input{psfig}

\begin{document}

\begin{flushright}
LU TP 04-28\\
October 18, 2004
\end{flushright}

\vspace{0.4in}

\begin{center}

{\LARGE \bf Folding Thermodynamics of Peptides} 

\vspace{.6in}

\large
Anders Irb\"ack and 
Sandipan Mohanty\footnote{E-mail: anders,\,sandipan@thep.lu.se}\\   
\vspace{0.10in}
Complex Systems Division, Department of Theoretical Physics\\ 
Lund University,  S\"olvegatan 14A,  SE-223 62 Lund, Sweden \\
{\tt http://www.thep.lu.se/complex/}\\

\vspace{0.3in}	

Submitted to \BJ

\end{center}
\vspace{0.6in}
\normalsize
Abstract:\\
A simplified interaction potential for protein folding studies
at the atomic level is discussed and tested
on a set of peptides with about 20 residues each. The test set contains
both $\alpha$-helical (Trp cage, \Fs) and $\beta$-sheet (GB1p, GB1m2,
GB1m3, Betanova, LLM) peptides. The model, which
is entirely sequence-based, is able to fold these different peptides
for one and the same choice of model parameters. Furthermore, the melting
behavior of the peptides is in good quantitative agreement with
experimental data. Apparent folded populations obtained using
different observables are compared, and are found to be very
different for some of the peptides (e.g., Betanova). In other
cases (in particular, GB1m2 and GB1m3), the different estimates
agree reasonably well, indicating a more two-state-like melting behavior.

\vspace{18pt}


\newpage 

\section{Introduction}

The function of peptides and proteins is inextricably connected to their
folding behavior, as is underlined by the facts that many neuro-degenerative
disorders are being linked to misfolding and aggregation~\cite{Dobson:03},
and that coupled folding and binding seems to be a
more common phenomenon than previously thought~\cite{Dyson:02}.
It is therefore an important development that folding simulations
at the atomic level are now becoming feasible for short
polypeptide chains~\cite{Gnanakaran:03}, thanks to faster computers,
more efficient algorithms and improved force fields.

There are, however, questions about the interaction
potentials used in the
simulations that need further investigation. One difficulty
is that different potentials give very different
relative weights to the $\alpha$-helix and $\beta$-strand regions of the
Ramachandran space~\cite{Zaman:03}. A potential that successfully
folds an $\alpha$-helical peptide might therefore have problems
with $\beta$-sheet peptides, and {\it vice versa}.
Another difficulty is with the temperature dependence of observable quantities.
As pointed out by Zhou et al.~\cite{Zhou:01}, 
it seems that most current models need
further calibration in order to give a temperature dependence that is
not too weak; as a result, calculated melting temperatures
are often unrealistically high. A systematic study of these thermodynamic
questions requires extensive conformational sampling
and is a challenge, especially in models with explicit water.

Here we study a model that contains all atoms of the polypeptide
chains but no explicit solvent molecules.
Formally, such a model is obtained by integrating out the
solvent degrees of freedom. Finding an accurate and computationally
tractable approximation of the resulting effective potential is, however,
a highly non-trivial problem.  Examples of implicit solvent models
that have been used in folding studies with some success,
include the generalized Born approach~\cite{Still:90},
the method based on screened Coulomb potentials 
by Hassan et al.~\cite{Hassan:03},
and the method based on solvent accessible surface areas
by Ferrara et al.~\cite{Ferrara:02}. In this paper, we study a minimalistic 
model in which the effects of the solvent are represented by an effective
attraction between nonpolar side chains. Our study focuses on the
thermodynamic behavior of this model, which we investigate using
efficient Monte Carlo methods rather than molecular dynamics.
This choice is made for computational convenience; with some minor
modifications, it would be possible to study the same model using
molecular dynamics.  Promising computational techniques have
recently been proposed by Hansmann and Wille~\cite{Hansmann:02} and
Schug et al.~\cite{Schug:03}, but these methods are for energy minimization,
which is insufficient for our purposes.

In addition to effective hydrophobic attraction,
the interaction potential of our model contains two major terms, representing
excluded-volume effects and hydrogen bonding.
The potential is deliberately kept simple, partly for the sake
of clarity but also for practical reasons; any potential requires careful
calibration, and this task is easier with a simple potential like ours with
fewer parameters to tune. In the future, the potential may be further
developed with the inclusion of new terms such as Coulomb interactions
between side-chain charges, but not before
it becomes clear that they are needed. The different terms of the potential
represent either the interaction between two individual atoms
(excluded volume), or two pairs of atoms (e.g., hydrogen bonds),
or an effective interaction between a pair of side chains (hydrophobicity).
The largest units playing a role in the potential are the amino acids,
and no information about the sequence as a whole or its native
structure is used in the potential.

Our approach towards the problem of determining the interaction potential
is phenomenological. The shape of individual terms is inspired by intuitive
notions rather than being rigorously derived from a microscopic picture.
Their exact functional forms and relative sizes are constrained by the
effectiveness of the model in describing the folding behavior of more and more
sequences. When such a potential evolves to a point where it can successfully
fold a significant number of peptides of different native geometries, and
capture the thermodynamic behavior of all those peptides, it would be useful
on its own as a working potential for thermodynamic studies of new sequences,
and also provide hints about the relative importance of different physical
effects in protein folding.

We have previously shown that earlier
versions of this model are able to fold both $\alpha$-helix and $\beta$-sheet
peptides~\cite{Irback:03,Irback:04}. In this paper
we present a further development of this model.
We test the new model on the following set of peptides (see Fig.~1):
the $\alpha$-helical Trp cage~\cite{Neidigh:02}
and \Fs~\cite{Lockhart:92,Lockhart:93}, and the $\beta$-sheet
peptides GB1p~\cite{Kobayashi:93,Blanco:94}, GB1m2 and
GB1m3~\cite{Fesinmeyer:04},
Betanova~\cite{Kortemme:98} and LLM~\cite{Lopez:01}.
Here GB1p denotes the C-terminal $\beta$-hairpin from the protein
G B1 domain, while Betanova is a designed three-stranded $\beta$-sheet
peptide. GB1m2 and GB1m3 are mutants of GB1p, while LLM is a mutant of
Betanova, with enhanced stabilities. We find that our model provides a good
description of the thermodynamic behavior of all these peptides.
The same model was furthermore used in a recent study of the oligomerization
properties of the amyloid \Aba\ peptide~\cite{Favrin:04},
with very promising results.

\begin{figure}[t]
\begin{center}
\epsfig{figure=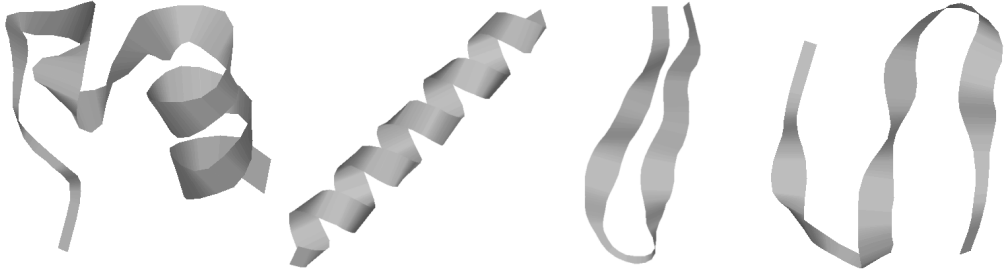,height=4.5cm}
\end{center}
\vspace{-3mm}
\caption{Schematic illustration of the different geometries of the 
peptides studied. 
Shown from left to right are the reference structures (see below) used for  
the Trp cage, \Fs, GB1m3 and Betanova. Drawn with 
RasMol~\protect\cite{Sayle:95}.}
\label{fig:1}\end{figure}

\section{Model and Methods}

\subsection{Model}

Our model contains all atoms
of the polypeptide chains, including hydrogen atoms.
The model assumes fixed bond lengths, bond angles and peptide
torsion angles ($180^\circ$),
so that each amino acid only has the Ramachandran torsion
angles $\phi$, $\psi$ and a number of side-chain torsion angles as its
degrees of freedom. Numerical values of the geometrical parameters
held constant can be found elsewhere~\cite{Irback:03}.

In the simulations we internally use a dimensionless energy scale.
The correspondence (constant factor) of this scale to the
physical energy scale is determined by using the model prediction of the
dimensionless energy  value for an observable and the experimental value
for the same. We use the melting temperature $\Tm=315$\,K of the
Trp cage~\cite{Neidigh:02} for this purpose (see below), which is
found to correspond to a dimensionless energy $k\Tm$ of 0.470 in the
model ($k$ is Boltzmann's constant). Energy parameters of the model
(such as the $\kev$, $\kloc$, $\ehba$, etc. below) are given in our
internal energy scale.  It must be emphasized that this energy scale is
left unchanged when analyzing the other peptides.

The interaction potential
\begin{equation}
  E=\Eev+\Eloc+\Ehb+\Ehp
\label{energy}
\end{equation}
is composed of four terms.
The first term in Eq.~\ref{energy}, $\Eev$, represents excluded-volume effects
and has the form
\beq
\Eev=\kev \sum_{i<j}
\biggl[\frac{\lambda_{ij}(\sigma_i+\sigma_j)}{r_{ij}}\biggr]^{12}\,,
\label{ev}\eeq
where the summation is over pairs of atoms $(i,j)$, $\kev=0.10$,
and $\sigma_i=1.77$, 1.75, 1.55, 1.42 and 1.00\,\AA\ for
S, C, N, O and H atoms, respectively. The values of the radii $\sigma_i$
agree reasonably well with the statistical analysis of
Tsai et al.~\cite{Tsai:99}. The $\sigma_i$ values for C, N and O
strongly influence the shape of the Ramachandran $\phi,\psi$ distribution,
and must therefore be carefully chosen. The parameter $\lambda_{ij}$ in
Eq.~\ref{ev} has the value 0.75 for all pairs except those connected by
three covalent bonds, for which $\lambda_{ij}=1$. The reason why
we use a reduction factor $\lambda_{ij}<1$ for all non-local
pairs is both computational efficiency and the restricted flexibility
of a chain with only torsional degrees of freedom, which could
create artificial traps. To speed up the calculations, Eq.~\ref{ev} is
evaluated using a cutoff of $\rc_{ij}=4.3\lambda_{ij}$\,\AA, and
pairs with fixed separation are omitted.

The second energy term, $\Eloc$, has the form
\beq
\Eloc=\kloc \sum_I
\left(\sum \frac{q_iq_j}{r_{ij}^{(I)}/{\rm \AA}}\right)\,,
\label{loc}\eeq
where the inner sum represents the interactions between
the partial charges of the backbone NH and \Cp O groups in one
amino acid, $I$. This potential is not used for Pro which lacks
the NH group, or Gly which tends to be more exposed to water
than other amino acids, due to the missing side chain. Neither is it used
for the two end amino acids, unless these are protected by
capping groups. The inner sum in Eq.~\ref{loc} has four terms
(NO, N\Cp, H\Cp\ and HO)
which depend only on the $\phi$ and $\psi$ angles for amino acid $I$.
The partial charges are taken as $q_i=\pm 0.20$ for H and N
and $q_i=\pm 0.42$ for \Cp\ and O~\cite{Branden:91}, and we put
$\kloc=100$,
corresponding to a dielectric constant of $\epsilon_r\approx 2.5$.

The third term of the energy function is the hydrogen-bond energy $\Ehb$,
which has the form
\begin{equation}
  \Ehb= \ehba \sum_{{\rm bb-bb}}
  u(r_{ij})v(\alpha_{ij},\beta_{ij}) +
  \ehbb \sum_{{\rm sc-bb}}
  u(r_{ij})v(\alpha_{ij},\beta_{ij})\,,
  \label{hbonds}
\end{equation}
where the two functions $u(r)$ and $v(\alpha,\beta)$ are given by
\begin{eqnarray}
u(r)&=& 5\bigg(\frac{\shb}{r}\bigg)^{12} -
             6\bigg(\frac{\shb}{r}\bigg)^{10}\label{u}\\
v(\alpha,\beta)&=&\left\{
        \begin{array}{ll}
             (\cos\alpha\cos\beta)^{1/2} &
              \ {\rm if}\ \alpha,\beta>90^{\circ}\label{v}\\
             0  & \ \mbox{otherwise}
        \end{array} \right.
\end{eqnarray}
We consider only hydrogen bonds between NH and CO groups, and
$r_{ij}$ denotes the HO distance, $\alpha_{ij}$ the NHO angle and
$\beta_{ij}$ the HOC angle. The parameters $\ehba$, $\ehbb$ and
$\shb$ are taken as 3.1, 2.0 and 2.0\,\AA, respectively.
The function $u(r)$ is calculated using a cutoff of $\rc=4.5$\,\AA.
The first sum in Eq.~\ref{hbonds} contains backbone-backbone
interactions, while the second sum contains interactions between
charged side chains (Asp, Glu, Lys and Arg) and the backbone.
The latter type of interaction is taken to be effectively weak
($\ehbb<\ehba$), because there are competing interactions between the
side-chain charges and the surrounding water that are omitted in the model.
For the same reason, we do not include any term in $\Ehb$ corresponding
to side chain-side chain interactions. It is possible that the effective
strength $\ehbb$ should be made stronger in case the side-chain charge
gets shielded from the water. This context dependence is ignored in the
model, which should be a reasonable approximation for small peptides.
Hydrogen bonds between parts that are very close in sequence are rare
in protein structures and therefore disregarded in the model;
specifically, we disallow backbone NH (\Cp O) groups to make
hydrogen bonds with the two nearest backbone \Cp O (NH) groups
on each side of them, and we also forbid hydrogen bonds
between the side chain of one amino acid
with the nearest donor or acceptor on either side of its \Ca. As a
simple form of context dependence, we assign a reduced strength to
hydrogen bonds involving chain ends, which tend to be exposed to water.
A hydrogen bond involving one or two end groups is reduced in strength
by factors of 2 and 4, respectively. If there are capping groups, these
groups are taken to be the end groups; otherwise, the two end amino
acids take this role.

The fourth energy term, $\Ehp$, represents an effective hydrophobic attraction
between nonpolar side chains. It has the pair-wise additive form
\beq
\Ehp=-\sum_{I<J}\Mij C_{IJ}\,,
\label{hp}\eeq
where $C_{IJ}$ is a measure of the degree of contact between
side chains $I$ and $J$, and $\Mij$ sets the energy that a pair
in full contact gets. The matrix $\Mij$ is defined in Table~1.
To calculate $C_{IJ}$ we use a predetermined set of atoms, $A_I$,
for each side chain $I$. We define $C_{IJ}$ as
\beq
C_{IJ}=\frac{1}{N_I+N_J} \biggl[\,
\sum_{i\in A_I}f(\min_{j\in A_J} r_{ij}^2) +
\sum_{j\in A_J}f(\min_{i\in A_I} r_{ij}^2)
\,\biggr]\,,
\label{Rij}\eeq
where the function $f(x)$ is given by $f(x)=1$ if $x<A$, $f(x)=0$ if $x>B$,
and $f(x)=(B-x)/(B-A)$ if $A<x<B$ [$A=(3.5\,{\rm \AA}){}^2$ and
$B=(4.5\,{\rm \AA}){}^2$]. Roughly speaking, $C_{IJ}$ is the fraction
of atoms in $A_I$ or $A_J$ that are in contact with some atom from
the other side chain. For Pro, the set $A_I$ consists of the
\Cb, \Cg\ and \Cd\ atoms. The definition of $A_I$ for the other
hydrophobic side chains has been given elsewhere~\cite{Irback:03}.
We expect the gain in forming a hydrophobic contact to be smaller
if the two side chains are close in sequence, because such a pair
is partly protected by the backbone. Therefore, we reduce the strength
of the hydrophobic attraction for pairs that are nearest
or next-nearest neighbors along the sequence; $\Mij$ is reduced by a factor
of 2 for next-nearest neighbors, and taken to be 0 for nearest neighbors.

\begin{table}[t]
\begin{center}
\begin{tabular}{rlccc}
                           & & I   & II & III \\
\hline
I& Ala                       & 0.0   & 0.1  & 0.1   \\
II& Ile, Leu, Met, Pro, Val  &       & 0.9  & 2.8   \\
III& Phe, Trp, Tyr           &       &      & 3.2
\end{tabular}
\caption{
The hydrophobicity matrix $\Mij$. Hydrophobic amino acids are divided
into three categories. The matrix $\Mij$ represents the size of
hydrophobicity interaction when an amino acid of type $I$ is in contact
with an amino acid of type $J$.}
\label{tab:1}
\end{center}
\end{table}

The parameters of this potential were essentially determined by a
somewhat tedious trial and error procedure, involving parallel
simulations of the different peptides. The target was to have
native-like free-energy minima for all the peptides at low temperature,
whereas the temperature dependence was not considered at all.
It is interesting to note that this criterion alone was sufficiently
discriminating to yield parameter values that appear physically
reasonable, as well as a realistic temperature dependence (see below).
Some parameters, such as $\ehba$, strongly influence the
folding properties of the model, and are therefore well determined.
Others, such as $\ehbb$, are less important and, as a result of this,
quite poorly determined.

The new version of the model differs from
earlier versions in the precise form of the simple context
dependence of $\Eloc$ and $\Ehb$. Also, the reduction factor for
the hydrophobic attraction between next-nearest neighbors
along the chain has been changed.
Furthermore, we have added Pro, which
does not occur in any of our previously studied sequences,
to the list of hydrophobic amino acids. All other parameters
of the potential are
the same as in the last version of the model, except for a slight reduction
in strength of the local potential ($\kloc$).

It should be stressed that this potential is not expected to provide a
good description of general amino acid sequences.
For example, it is likely that the pair-wise additive hydrophobicity
potential is inadequate for long chains, due to double-counting effects.
For long chains, anti-cooperative multibody effects might play
a significant role~\cite{Shimizu:01}. By extending the present
calculations in the future to new and longer sequences, we hope
that it will be possible to refine the potential and
thereby make it more general.

\subsection{Computational methods}

To study the thermodynamic behavior of this model,
we use simulated 
tempering~\cite{Lyubartsev:92,Marinari:92,Irback:95}.
in which the temperature is a dynamical variable.
For a review of simulated tempering and other generalized-ensemble
techniques for protein folding, see Hansmann and 
Okamoto~\cite{Hansmann:99}. We study
eight different temperatures $T_k$,
which range from $\Tmin=275$\,K to $\Tmax=369$\,K and
are given by $T_k=\Tmin(\Tmax/\Tmin)^{(k-1)/7}$ ($k=1,\ldots,8$).
The average acceptance rate for the temperature jumps
is about 70\,\%.

Our simulations are carried out using two different elementary moves
for the backbone degrees of freedom:
first, the highly non-local pivot move in which a single backbone
torsion angle is turned; and second, a semi-local
method~\cite{Favrin:01} that
works with up to eight adjacent backbone degrees of freedom, which are turned
in a coordinated manner. Side-chain angles are updated one by one.
Every update involves a Metropolis accept/reject step,
thus ensuring detailed balance. All our simulations are started from
random configurations. All statistical errors quoted are 1$\sigma$ errors
obtained from the variation between independent runs. For each peptide,
we performed about 10 independent runs. Each run contained $10^9$ elementary
Monte Carlo steps ($1.5\cdot10^9$ steps for GB1p) and required
1--2 CPU days on a 1.6\,GHz computer.

To characterize the folding behavior of the different peptides, we monitor
several quantities. For a peptide with $N$ amino acids, we define the
$\alpha$-helix content $H$ as the fraction of the $N-2$
inner amino acids with their Ramachandran ($\phi,\psi$) pair in the
region $-90^\circ<\phi<-30^\circ$, $-77^\circ<\psi<-17^\circ$.
We calculate the radius of gyration, $\Rg$, over the backbone
atoms, with unit mass for all atoms. We also study root-mean-square
deviations (RMSD) from folded reference structures, calculated over either
the backbone atoms or all heavy atoms. A backbone RMSD is denoted by $\Db$
and a heavy-atom RMSD by $\Delta$.
For the $\beta$-sheet peptides, there exist topologically distinct
states that the backbone RMSD cannot discriminate between, which
makes it necessary to use the heavy-atom RMSD.

In our analysis of the results from the simulations, it turns out that
the temperature dependence of a quantity $X$ in many cases can be
well described by the simple two-state expression
\beq
X(T)=\frac{\Xu+\Xn K(T)}{1+K(T)}\,.
\label{twostate}\eeq
Our fits to this equation are carried out by using a Levenberg-Marquardt
procedure~\cite{NR}.
Throughout the paper, the baselines $\Xu$ and $\Xn$
are taken to be temperature independent, whereas the effective equilibrium
constant $K(T)$ is assumed to have the first-order
form $K(T)=\exp[(1/kT-1/k\Tm)\dE]$, where $\Tm$ is
the midpoint temperature and $\dE=\Eu-\En$ is the energy difference
between the unfolded and native states.
With these assumptions, a fit to Eq.~\ref{twostate} has four
parameters: $\dE$, $\Tm$, $\Xu$ and $\Xn$.

\section{Results and Discussion}

Using the model and methods described in the previous section, we
performed high-statistics thermodynamic simulations of the
peptides mentioned in the introduction, namely the Trp cage, \Fs,
GB1p, GB1m2, GB1m3, Betanova and LLM. In this section we
present the results of these calculations.

\subsection{Trp cage}

The optimized 20-residue Trp cage (NLYIQWLKDGGPSSGRPPPS) is a
``miniprotein'' with a compact folded state and a melting
temperature of 315\,K, as determined by circular dichroism (CD)
and NMR measurements~\cite{Neidigh:02}.
The NMR-derived native structure~\cite{Neidigh:02}
contains a short $\alpha$-helix (residues 2--8), a single turn of
$3_{10}$-helix (residues 11-14), and a hydrophobic core consisting
of three proline residues (Pro12, Pro18, Pro19) and two
aromatic residues (Tyr3, Trp6). The folding time is
a few $\mu$s at room temperature~\cite{Qiu:02}. Its small size, fast
folding and relative stability makes the Trp cage an ideal testbed for
computational methods, and folding simulations of this peptide were reported
by several groups~\cite{Snow:02,Simmerling:02,Schug:03,Pitera:03,Zhou:03a}.
Two of these groups performed thermodynamic
studies~\cite{Pitera:03,Zhou:03a}.
Both groups made detailed comparisons with raw NMR data with very
good results, but the calculated melting temperatures
were too high ($\gtrsim 400$\,K).

In our model the melting temperature of the Trp cage is, by definition,
equal to its experimental value, since we use this quantity to set
the energy scale of the model. For this purpose, we consider the helix
content $H$, as defined in the previous section, which should
be strongly correlated with the CD signal studied experimentally.
Fig.~2a shows our results for $H$ against temperature.
\begin{figure}[t]
\begin{center}
\epsfig{figure=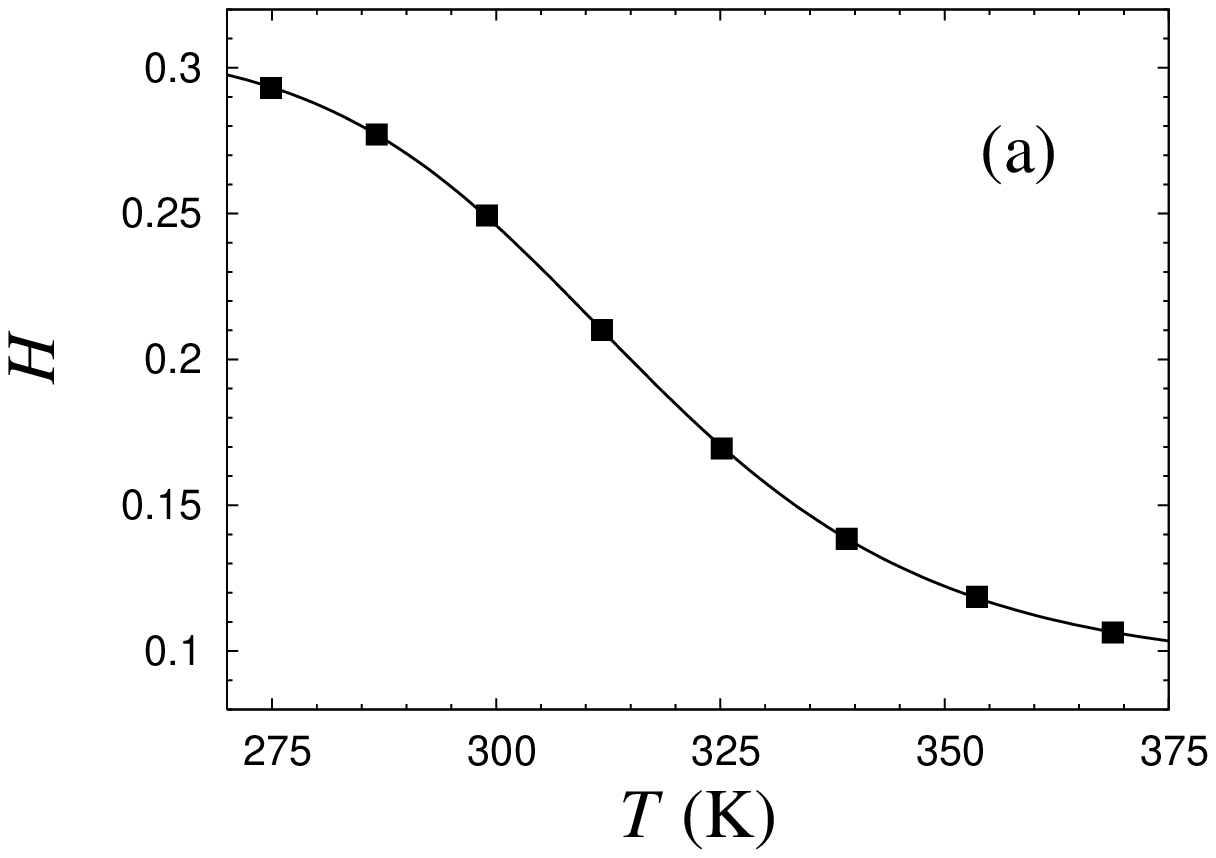,width=7cm}
\epsfig{figure=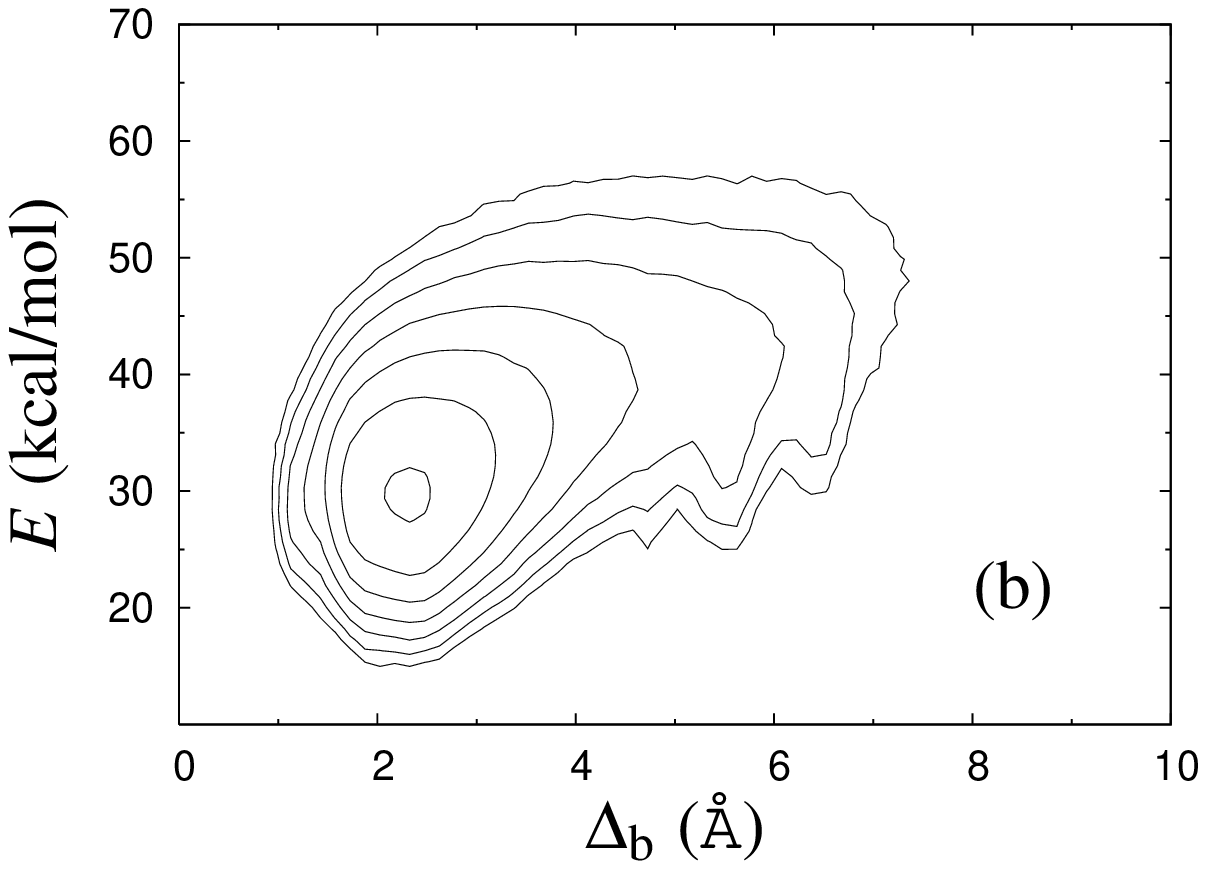,width=7cm}
\end{center}
\vspace{-3mm}
\caption{The Trp cage. (a) Helix content against temperature. 
The line is a fit to Eq.~\protect\ref{twostate}
($\Tm=315$\,K, $\dE=11.5\pm0.2$\,kcal/mol). 
Statistical errors are smaller than the plot symbols.
(b) Contour plot of the free energy $F(\Db,E)$ at 275\,K. 
The contours are spaced at intervals of 1\,$kT$. Contours more than
6\,$kT$ above the minimum free energy are not shown. The free energy
 $F(\Db,E)$ is defined by $\exp[-F(\Db,E)/kT]\propto P(\Db,E)$, where 
$P(\Db,E)$ denotes the joint probability distribution of $\Db$ and $E$ 
at temperature $T$.}
\label{fig:2}\end{figure}
A fit to the data with the two-state expression in
Eq.~\ref{twostate} is also shown. As can be seen in the figure,  the two-state
fit provides an excellent description of the
data. The midpoint temperature from this fit, $\Tm$, is set to 315\,K,
the experimental melting temperature. Having done that,
there is no free parameter left in the model. The
fitted value of the parameter $\dE=11.5\pm0.2$\,kcal/mol is, in
contrast to that of $\Tm$, not used for calibration, but is rather
a prediction of the model.

In the two-state picture (Eq.~\ref{twostate}), the native population at
temperature $T$ is given by $1/\{1+\exp[-(1/kT-1/k\Tm)\dE]\}$.
Fig.~3 shows the native
population obtained using the above mentioned $\dE$~ and $\Tm$, against
temperature, along with
experimental values based on CD and NMR~\cite{Neidigh:02}.
\begin{figure}[t]
\begin{center}
\epsfig{figure=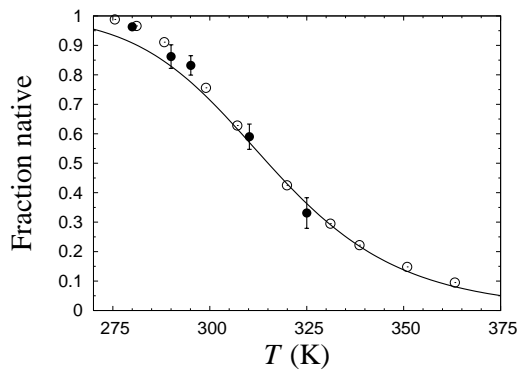,width=7cm}
\end{center}
\vspace{-3mm}
\caption{Native population against temperature for the Trp cage.
The line is the result obtained from the model, through the fit  
shown in Fig.~\protect\ref{fig:2}a. Plot symbols show experimental 
results~\cite{Neidigh:02} based on CD ($\circ$) and NMR ($\bullet$), 
respectively.}  
\label{fig:3}\end{figure}
We see that the results obtained from the model are in good agreement
with the experimental data over the entire temperature range, with a
maximum deviation of $\sim$\,5\,\% at the lowest temperatures. With the
overall energy scale properly determined, we thus find that the melting
behavior of this peptide is well described by the model.

At low temperature, we find a helix content similar to that of
the NMR structure, $\sim$\,30\,\% (see Fig.~2a). An RMSD analysis
confirms that the typical low-temperature structure
is similar to the NMR structure (PDB code 1L2Y, first model),
as illustrated in  Fig.~2b.
This figure shows the free energy $F(\Db,E)$ calculated as a function
of the backbone RMSD $\Db$ (residues 2--19)
and the energy $E$, at $275$\,K. We see that $F(\Db,E)$ has a
simple shape with one dominating minimum,
which is located at $\Db\approx 2.3$\,\AA .

\subsection{\Fs}

The designed 21-residue \Fs\ peptide is given by 
Suc-A${}_5$(AAARA)${}_3$A-NH${}_2$,
(where Suc is succinylic acid) and makes an
$\alpha$-helix~\cite{Lockhart:92,Lockhart:93}. Other N-capping
groups than Suc have also been used in the experiments on this
peptide. The melting behavior of \Fs\ was studied using CD as well as
infrared (IR) spectroscopy. The melting temperature measured by IR was
334\,K~\cite{Williams:96}, whereas the CD-based studies obtained
$\Tm=308$\,K~\cite{Lockhart:93} and $\Tm=303$\,K~\cite{Thompson:97}.
Computational studies of \Fs\ have also been
reported~\cite{Vila:00,Garcia:02,Nymeyer:03}. 
By explicit water simulations,
Garc\'\i a and Sanbonmatsu~\cite{Garcia:02}
obtained a $\Tm$ of 345\,K,
which is in reasonable agreement with the IR-based value.
Using an earlier version of our model and ignoring the capping groups,
a $\Tm$ of 310\,K was obtained~\cite{Irback:03}.
In the present calculations, we include the Suc and NH${}_2$ groups.

Fig.~4a shows the helix content versus temperature
as obtained from our \Fs\ calculations.
\begin{figure}[t]
\begin{center}
\epsfig{figure=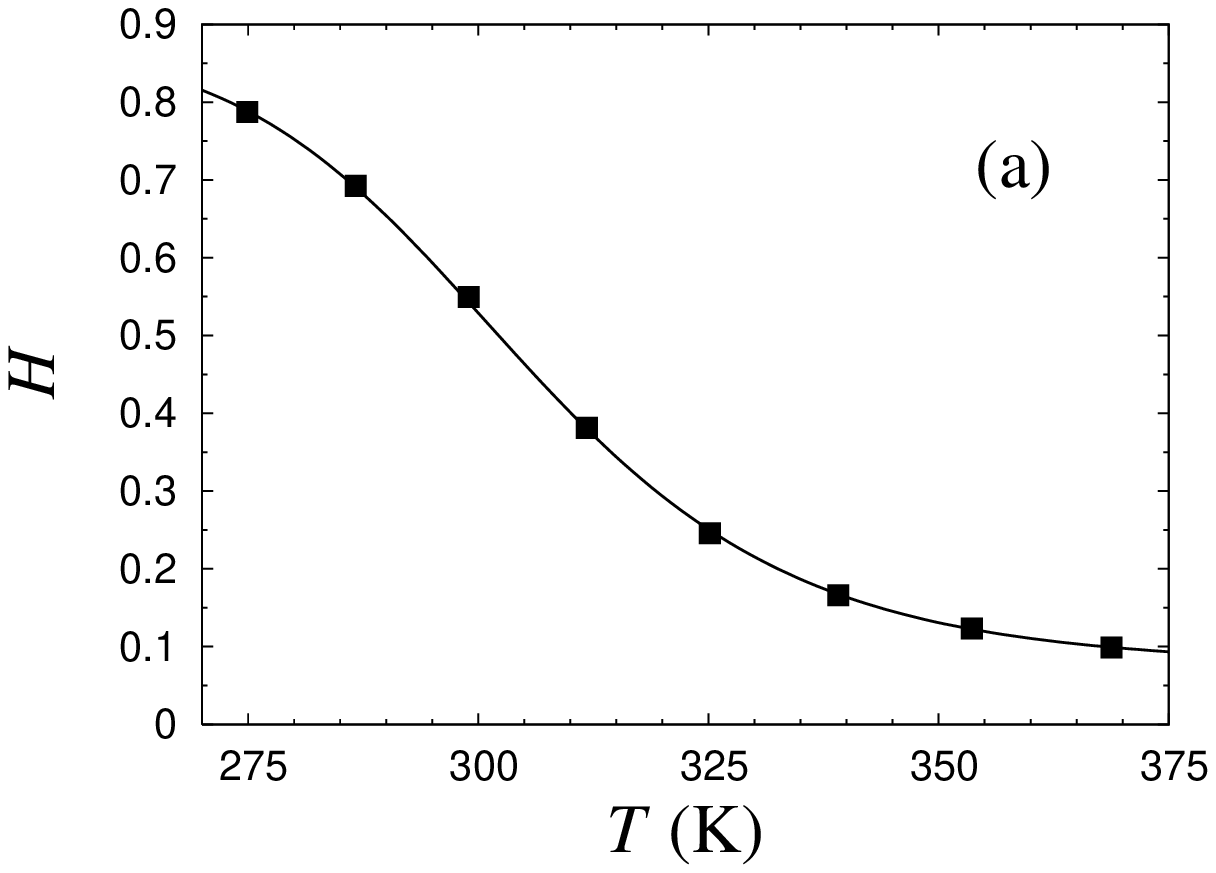,width=7cm}
\epsfig{figure=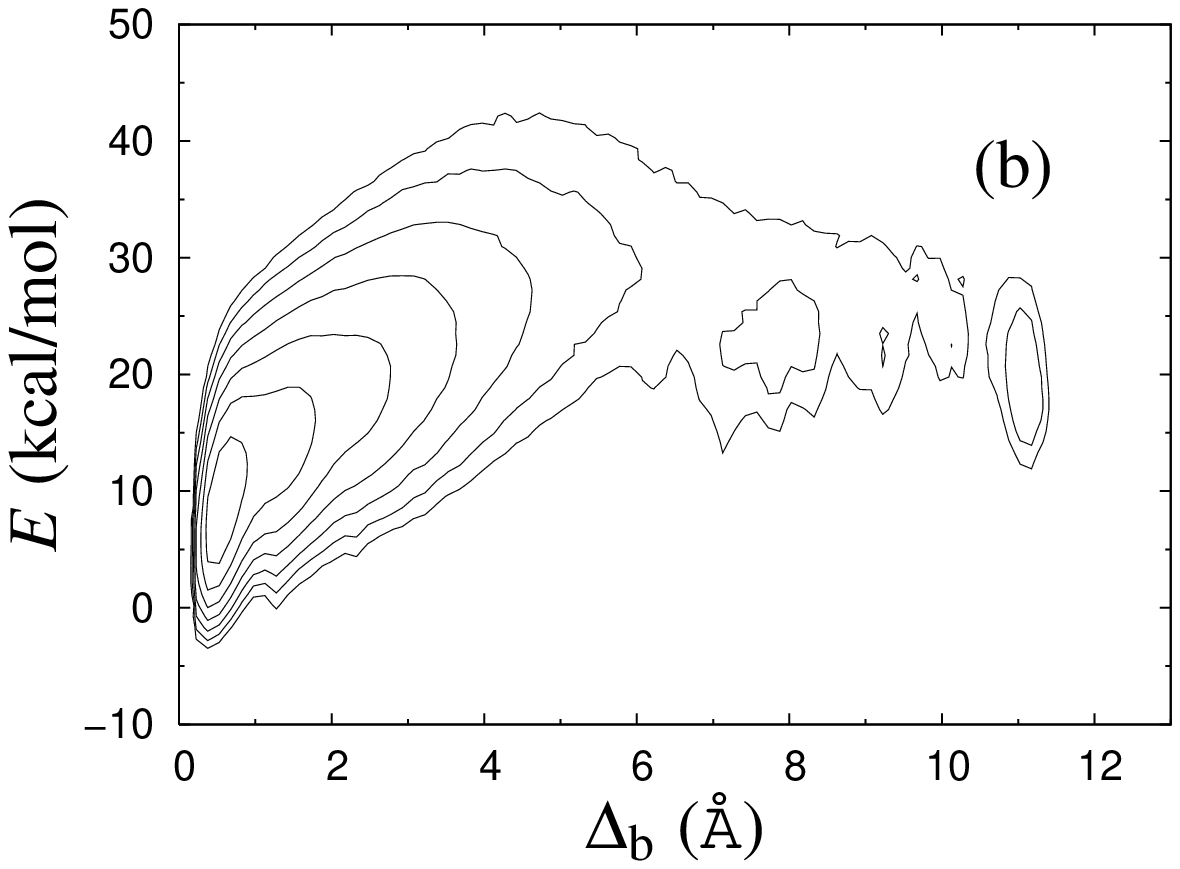,width=7cm}
\end{center}
\vspace{-3mm}
\caption{Same as Fig.~\protect\ref{fig:2} for the \Fs\ peptide 
($\Tm=304\pm1$\,K, $\dE=11.9\pm0.3$\,kcal/mol).}
\label{fig:4}\end{figure}
A two-state fit of the data gives $\Tm=304\pm1$\,K,
which is significantly lower than the IR-based
result mentioned above but in perfect agreement with the CD studies,
especially that of Thompson et al.~\cite{Thompson:97}. For
the energy difference, we obtain $\dE=11.9\pm0.3$\,kcal/mol,
which also agrees with what Thompson et al. found, namely
$\dE=12\pm 2$\,kcal/mol. It may be worth noting that the experimental
data that we compared with in the Trp cage case were based on CD rather
than IR.

In Fig.~4b we show the free energy $F(\Db,E)$ at
275\,K. In the absence of a precise experimental structure for \Fs, we define
$\Db$ as the (backbone) RMSD from an ideal $\alpha$-helix (all residues).
From the figure we see that the free energy has its global minimum
at $\Db\approx 0.5$\,\AA, which indeed corresponds to the $\alpha$-helix. There
are also two local minima at $\Db\approx 7$\,\AA\ and $\Db\approx 11$\,\AA,
both of which correspond to $\beta$-sheet structures. These two minima are
very weakly populated compared to the $\alpha$-helix minimum.

\subsection{GB1p and GB1m2/GB1m3}

Using exactly the same model, we now turn to $\beta$-sheet peptides.
That GB1p (GEWTYDDATKTFTVTE),
the 41--56-residue fragment from the protein G B1 domain,
makes a $\beta$-hairpin on its own was a breakthrough
discovery~\cite{Blanco:94} that has been followed by numerous
atomic simulations of this particular
sequence~\cite{Roccatano:99,Pande:99,Dinner:99,Garcia:01,
Zhou:01,Zagrovic:01,Kussell:02,Zhou:03b,Bolhuis:03,Wei:04}.
Recently, two mutants of GB1p with enhanced
stability were designed~\cite{Fesinmeyer:04}, GB1m2 and GB1m3,
by replacing the turn segment DDATKT by NPATGK. The mutant GB1m2
(GEWTYNPATGKFTVTE) is identical to GB1p except for this change,
while GB1m3 (KKWTYNPATGKFTVQE) differs from GB1p at the chain ends as well.
By CD and NMR, GB1m3 was estimated to be $86\pm3$\% folded
at 298\,K and to have a $\Tm$ of $333\pm2$\,K, whereas GB1m2 was
found to have a slightly lower folded population, $74\pm5$\,\% at 298\,K,
and a $\Tm$ of $320\pm2$\,K~\cite{Fesinmeyer:04}.
In the same study,
GB1p was estimated to be $\sim$\,30\,\% folded at 298\,K. An earlier NMR
study found GB1p to be 42\,\% folded at 278\,K~\cite{Blanco:94}. Both
these estimates of native population for GB1p are low compared to
the result of a Trp fluorescence study~\cite{Munoz:97}; a two-state
analysis of these data gave $\Tm=297$\,K and
$\dE=11.6$\,kcal/mol~\cite{Munoz:97}.

It turns out that our model fails to reproduce the experimental
difference in stability between GB1m2 and GB1m3. In fact,
GB1m2 and GB1m3 show nearly identical behavior in our model. For clarity,
we therefore show results only for one of these peptides, GB1m3, in the
figures below.

Fig.~5 shows the hydrophobicity energy $\Ehp$ against
temperature for GB1p and GB1m3 in the model.
\begin{figure}[t]
\begin{center}
\epsfig{figure=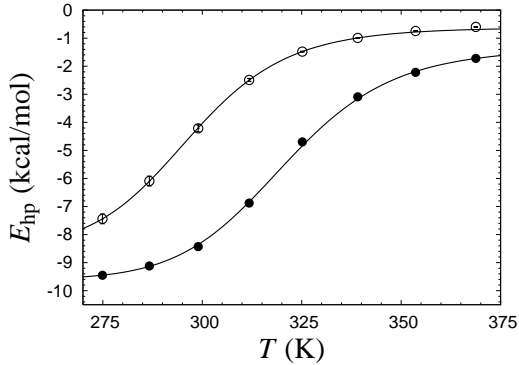,width=7cm}
\end{center}
\vspace{-3mm}
\caption{The hydrophobicity energy $\Ehp$ against
temperature for GB1p ($\circ$) and GB1m3 ($\bullet$). 
The lines are fits to Eq.~\protect\ref{twostate} 
($\Tm=297\pm1$\,K, $\dE=14.2\pm0.2$\,kcal/mol for GB1p; 
$\Tm=321\pm1$\,K, $\dE=15.0\pm0.4$\,kcal/mol for GB1m3). The 
points corresponding to the two highest temperatures were omitted
for GB1p, as removing them resulted in a significantly better fit in terms
of $\chi^2$ per degree of freedom.}
\label{fig:5}\end{figure}
We expect $\Ehp$ to be strongly correlated with Trp fluorescence
for these peptides, as Trp43 forms a hydrophobic
cluster together with Tyr45, Phe52 and Val54. A two-state fit to our
data for GB1p gives $\Tm=297\pm1$\,K and $\dE=14.2\pm0.2$\,kcal/mol, which
indeed is in good agreement with the Trp fluorescence
results for this peptide ($\Tm=297$\,K, $\dE=11.6$\,kcal/mol).
The same type of fit gives $\Tm=321\pm1$\,K and
$\dE=15.0\pm0.4$\,kcal/mol for GB1m3, and $\Tm=322\pm2$\,K and
$\dE=15.1\pm0.4$\,kcal/mol for GB1m2.
These two very similar $\Tm$ estimates lie close to the experimental
result for GB1m2 ($320\pm2$\,K) and
somewhat below that for GB1m3 ($333\pm2$\,K).
Our $\Ehp$ data indicate that GB1m2 and GB1m3 indeed are markedly
more stable than GB1p in the model, which is confirmed by the
results discussed next.

Fig.~6a shows our data for the free energy $F(\Delta,E)$
for GB1p, at 275\,K. 
\begin{figure}[t]
\begin{center}
\epsfig{figure=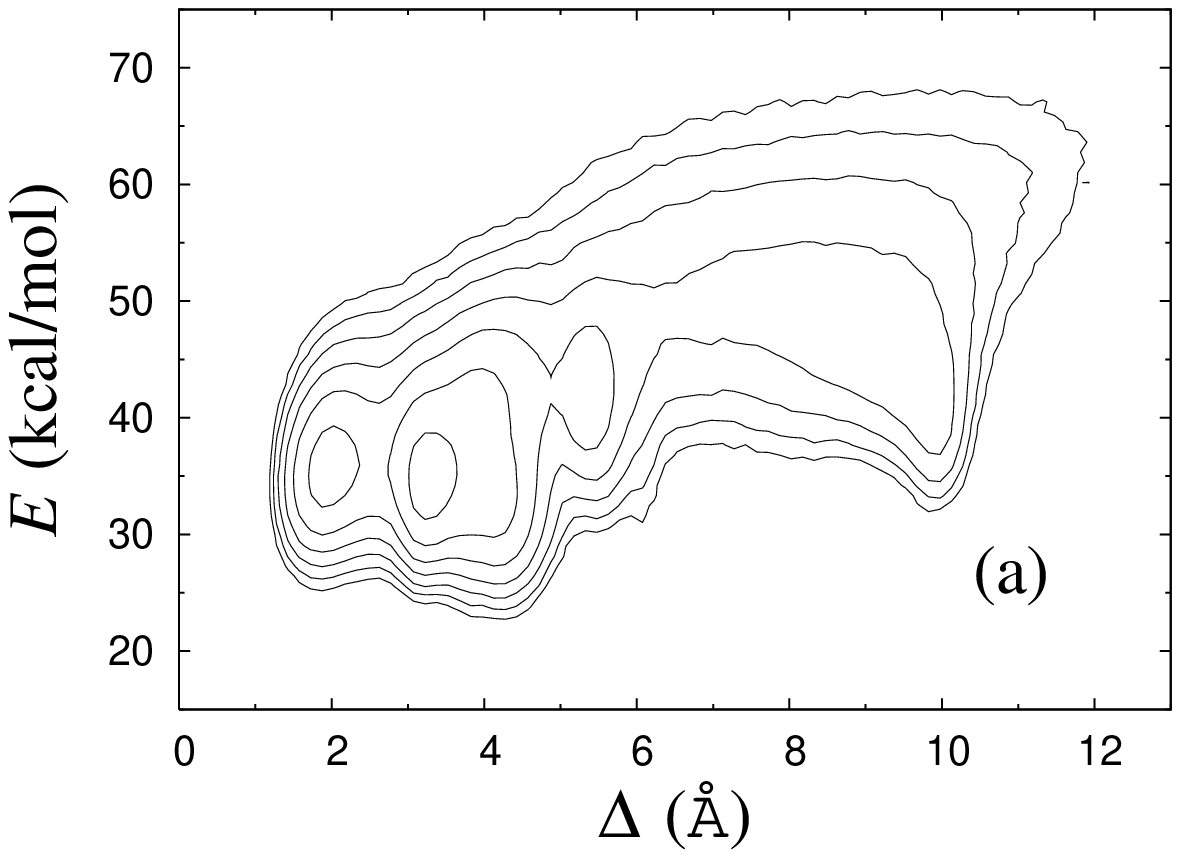,width=7cm}
\epsfig{figure=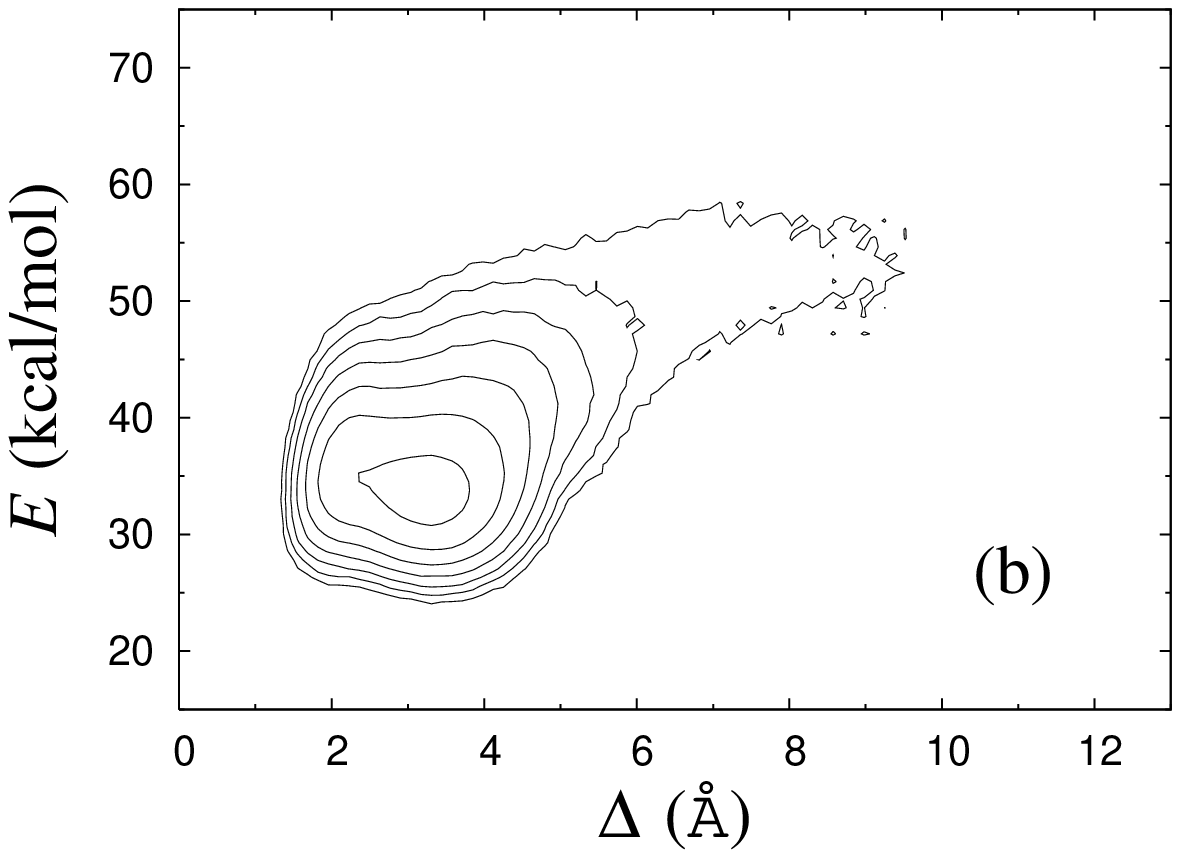,width=7cm}
\end{center}
\vspace{-3mm}
\caption{Contour plot of the free energy $F(\Delta,E)$ for (a) GB1p and
(b) GB1m3, at 275\,K. 
Contour levels are as in Fig.~\protect\ref{fig:2}b.}
\label{fig:6}\end{figure}
On its own the GB1p fragment is
believed to adopt a folded structure
similar to that it has as part of the native protein G B1 domain, although
the NMR restraints were insufficient to determine a unique structure
for the excised fragment.
As reference structure in the calculation of $\Delta$, we therefore use
the corresponding fragment of the NMR structure for the
full protein G B1 domain (PDB code 1GB1, residues 41--56,
first model)~\cite{Gronenborn:91}.
The heavy-atom RMSD $\Delta$ is used instead of
the backbone RMSD $\Db$, because $\Db$ cannot
distinguish between the two possible $\beta$-hairpin
topologies (with similar backbone folds but oppositely oriented
side chains). We find that the two lowest minima of $F(\Delta,E)$,
at $\Delta\approx2.0$\,\AA\ and $\Delta\approx3.2$\,\AA, both
correspond to a $\beta$-hairpin with the same topology and the
same set of backbone hydrogen bonds as the reference structure.
The main difference between these two minima lies in the shape of the turn
region. In addition to these minima, there are two weakly populated
local minima at $\Delta\approx5.3$\,\AA\ and $\Delta\approx8$--10\,\AA,
which correspond to a $\beta$-hairpin with the opposite topology and
$\alpha$-helix, respectively. The shape of $F(\Delta,E)$ for GB1p
was also studied using earlier versions of our
model~\cite{Irback:03,Irback:04}.
The present model yields very
similar results, with a minor enhancement of the two native-like minima
at the expense of the two other local minima mentioned above.

Fig.~6b shows the corresponding free-energy plot for GB1m3.
As reference structure for GB1m3, we use a
mutated and relaxed version of the GB1p reference structure.
We see that $F(\Delta,E)$ has a simpler shape for GB1m3 than for GB1p.
There is only one detectable free-energy minimum for GB1m3,
and this minimum corresponds to a
structure similar to the favored one for GB1p.

Different experiments on GB1p have, as mentioned above, obtained different
$\beta$-hairpin populations. One way of estimating folded populations
in the model is by two-state fits like those in Fig.~5.
An independent and more direct estimate can be
obtained by counting native backbone hydrogen bonds.
To this end, we consider a hydrogen bond
formed if its energy is less than $-\ehba/3$. The number of
native backbone hydrogen bonds in a given conformation is denoted
by $\Nhb$. Fig.~7 shows the probability
distribution of $\Nhb$ for GB1p and GB1m3 at 299\,K, which is very close
to the temperature (298\,K) at which the folded populations of these
two peptides were compared by CD and NMR~\cite{Fesinmeyer:04}.
\begin{figure}[t]
\begin{center}
\epsfig{figure=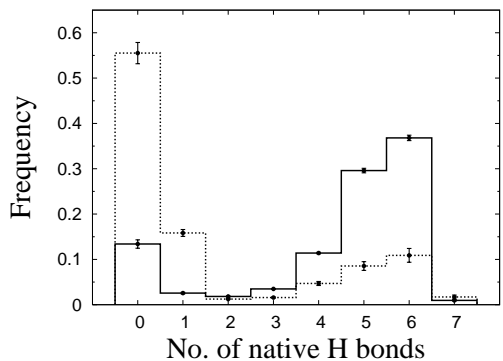,width=7cm}
\end{center}
\vspace{-3mm}
\caption{Probability distribution of the number of native  
hydrogen bonds, $\Nhb$, for GB1m3 (full line) and  GB1p (dotted line) 
at 299\,K. The hydrogen bonds taken as native are the same for both
peptides. In GB1p notation, the native hydrogen bonds are Glu42(N)-Thr55(O),
Glu42(O)-Thr55(N), Thr44(N)-Thr53(O), Thr44(O)-Thr53(N), 
Asp46(N)-Thr51(O), Asp46(O)-Thr51(N) and Asp47(O)-Lys50(N).}
\label{fig:7}\end{figure}
We find that the probability distribution
$P(\Nhb)$ has a clear bimodal shape for both peptides, with one
native and one unfolded peak. The native peak is, as expected from
the results above, significantly larger for the mutant GB1m3 than for GB1p.
Taking conformations with $\Nhb\ge 3$ as native and those with
$\Nhb\le 2$ as unfolded, we obtain native populations of
$82\pm1$\,\% for GB1m3, $84\pm1$\,\% for GB1m2, and $27\pm2$\,\%
for GB1p. The overall agreement between these results and the
experimental data ($86\pm3$\,\% for GB1m3, $74\pm5$\,\% for GB1m2,
$\sim$\,30\,\% for GB1p) is very good, although the model
slightly overestimates the folded fraction for GB1m2.
Note that the native
populations estimated from $P(\Nhb)$, thanks to the bimodality, are
quite well determined, despite that the precise definition of
native in terms of $\Nhb$ is somewhat arbitrary.

For GB1m3, we find that one of the hydrogen bonds taken as native
is very unlikely to form in our model, namely Pro47(O)-Gly50(N).
As a result, conformations with $\Nhb=7$ are very rare (see Fig.~7).

Our $\Ehp$- and $\Nhb$-based native populations for GB1p
are different; from the $\Ehp$ data we obtain
a native population of 46\,\% at 299\,K, where the $\Nhb$ analysis
gives 27\,\%. The magnitude of this difference
is similar to that between different experiments. The $\Nhb$-based
result is is good agreement with CD and NMR data, whereas the
$\Ehp$-based result agrees with Trp fluorescence data. For GB1m3
(and GB1m2), we do not know of any Trp fluorescence study. Our model
suggests that the difference between different methods would be
smaller in this case. Our $\Ehp$-based folded population at 299\,K
is 85\,\% for GB1m3, which is close to our $\Nhb$-based result of 82\,\%.

\subsection{Betanova and LLM}

Betanova is a designed antiparallel three-stranded $\beta$-sheet peptide with
20 residues (RGWSVQNGKYTNNGKTTEGR)~\cite{Kortemme:98}, which is only
marginally stable~\cite{Lopez:01}. Recently, Betanova
mutants with higher stability were developed~\cite{Lopez:01},
such as the triple mutant LLM (Val5Leu, Asn12Leu, Thr17Met).
The NMR-based native populations
of LLM and Betanova are 36\,\% and 9\,\%, respectively,
at 283\,K~\cite{Lopez:01}. Results in good agreement with
these estimates were obtained when testing an earlier version of our model
on these two peptides~\cite{Irback:04}. Folding simulations of
Betanova have also been performed by other groups, using
coarse-grained~\cite{Kim:04} and
atomic~\cite{Bursulaya:99,Colombo:02} models.

The folded structure of Betanova and LLM contains
eight backbone hydrogen bonds, four in each of the two
$\beta$-hairpins. Fig.~8a shows the
probability distribution of the number of native backbone hydrogen
bonds, $\Nhb$, in our model for LLM and Betanova, at 287\,K.
\begin{figure}[t]
\begin{center}
\epsfig{figure=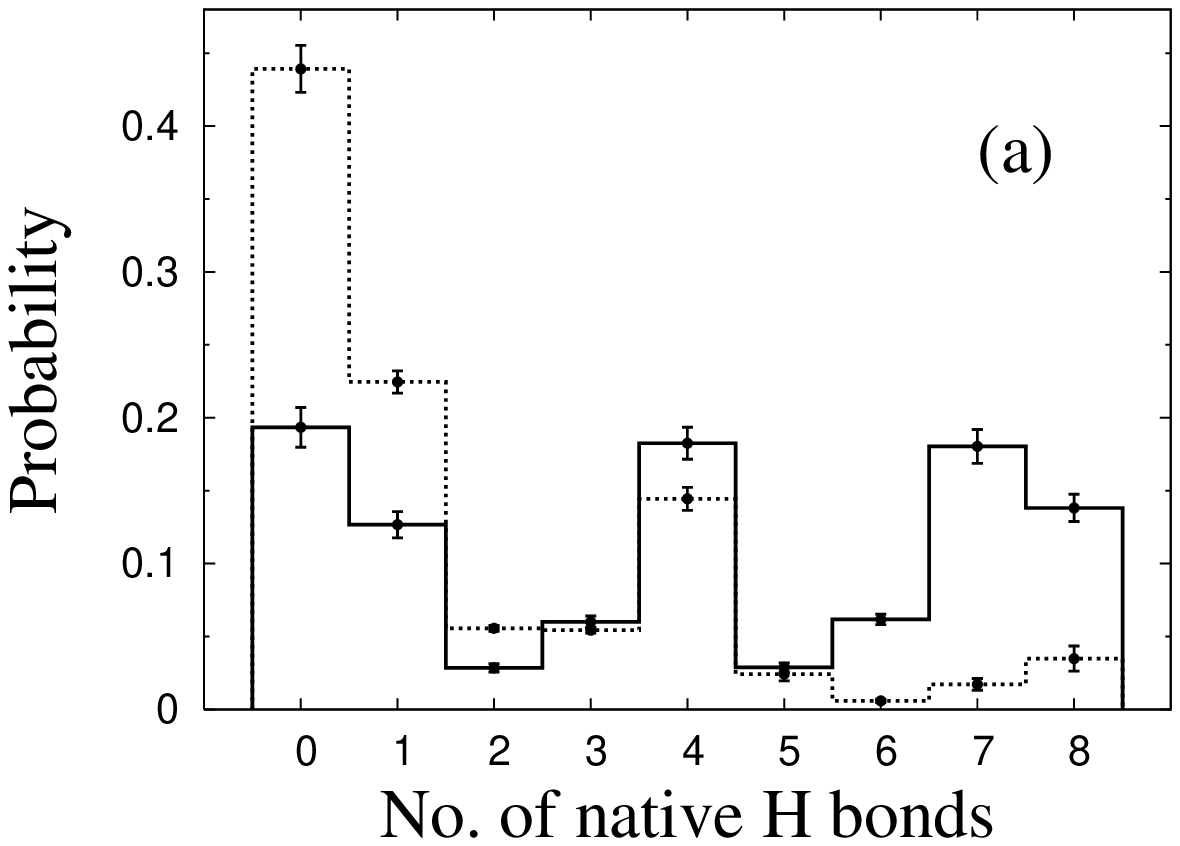,width=7cm}
\epsfig{figure=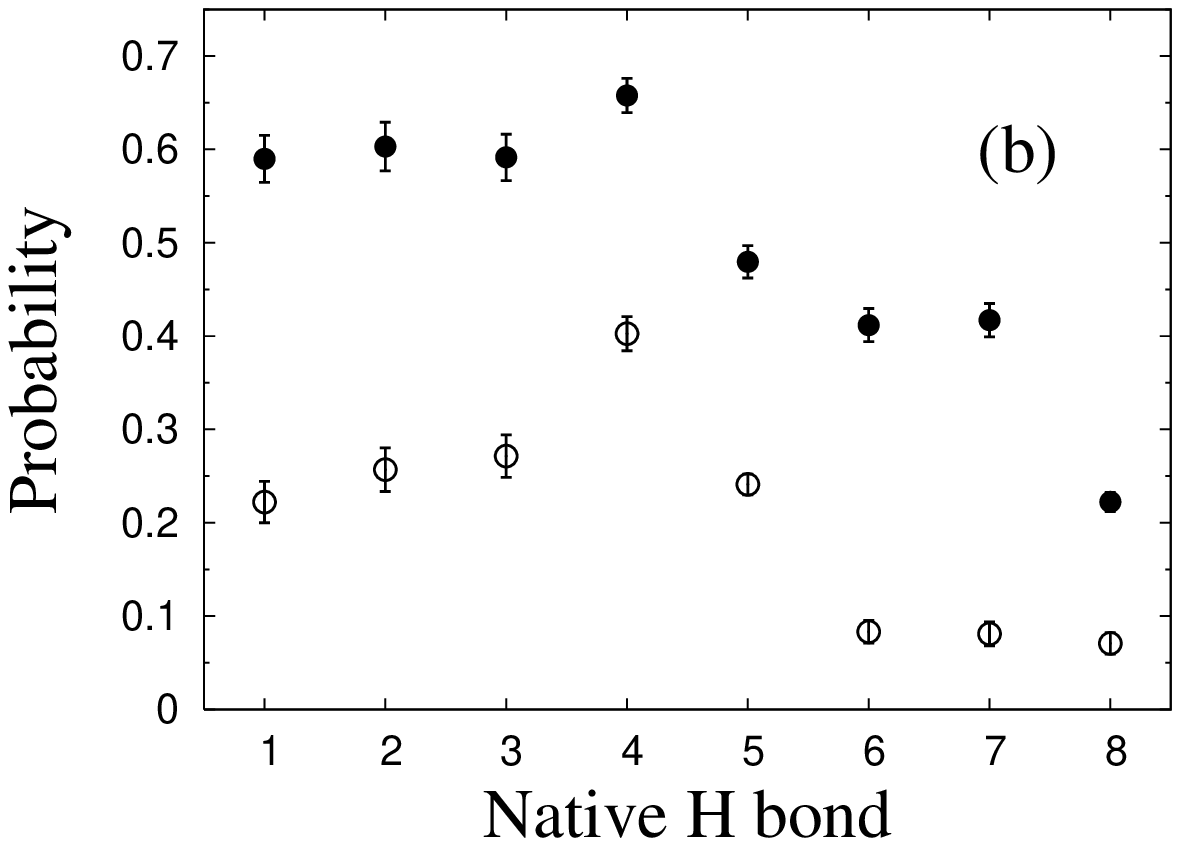,width=7cm}
\end{center}
\vspace{-3mm}
\caption{(a) Probability distribution of the number of native backbone
hydrogen bonds, $\Nhb$, for LLM (full line) and Betanova (dotted line)
at 287\,K. (b) Frequencies of occurrence for the different native
hydrogen bonds for Betanova ($\circ$) and LLM ($\bullet$) at 287\,K. 
In Betanova notation, the native hydrogen bonds are
1: Ser4(N)-Thr11(O), 2: Ser4(O)-Thr11(N), 3: Gln6(N)-Lys9(O), 
4: Gln6(O)-Lys9(N), 5: Tyr10(N)-Thr17(O), 6: Tyr10(O)-Thr17(N), 
7: Asn12(N)-Lys15(O) and 8: Asn12(O)-Lys15(N).} 
\label{fig:8}\end{figure}
The distributions have three peaks. In addition to the folded and
unfolded peaks at high and low $\Nhb$, there is also a peak at
$\Nhb=4$. Visual inspection of snapshots from the simulations
reveals that conformations at this peak tend to contain the
first (N-terminal) $\beta$-hairpin but not the second (C-terminal) one.
This conclusion, which is in agreement with experimental
data~\cite{Lopez:01},
is confirmed by the frequencies of occurrence of the
individual hydrogen bonds, shown in Fig.~8b.
We see that the hydrogen bonds of the first $\beta$-hairpin (1--4)
occur more frequently than those of the second $\beta$-hairpin (5--8),
especially for Betanova. For a conformation to be counted as folded,
we require that $\Nhb\ge6$. With this definition, we find that
Betanova and LLM are $6\pm1$\,\% and $38\pm2$\,\% folded, respectively, at
287\,K, which is in good agreement with the experimental
results (9\,\% and 36\% at 283\,K).

The melting behavior has, as far as we know, not been
studied experimentally for Betanova or LLM. In Fig.~9a
we show melting curves for these peptides in our model.
As in the $\beta$-hairpin case, we consider the hydrophobicity
energy $\Ehp$.
\begin{figure}[t]
\begin{center}
\epsfig{figure=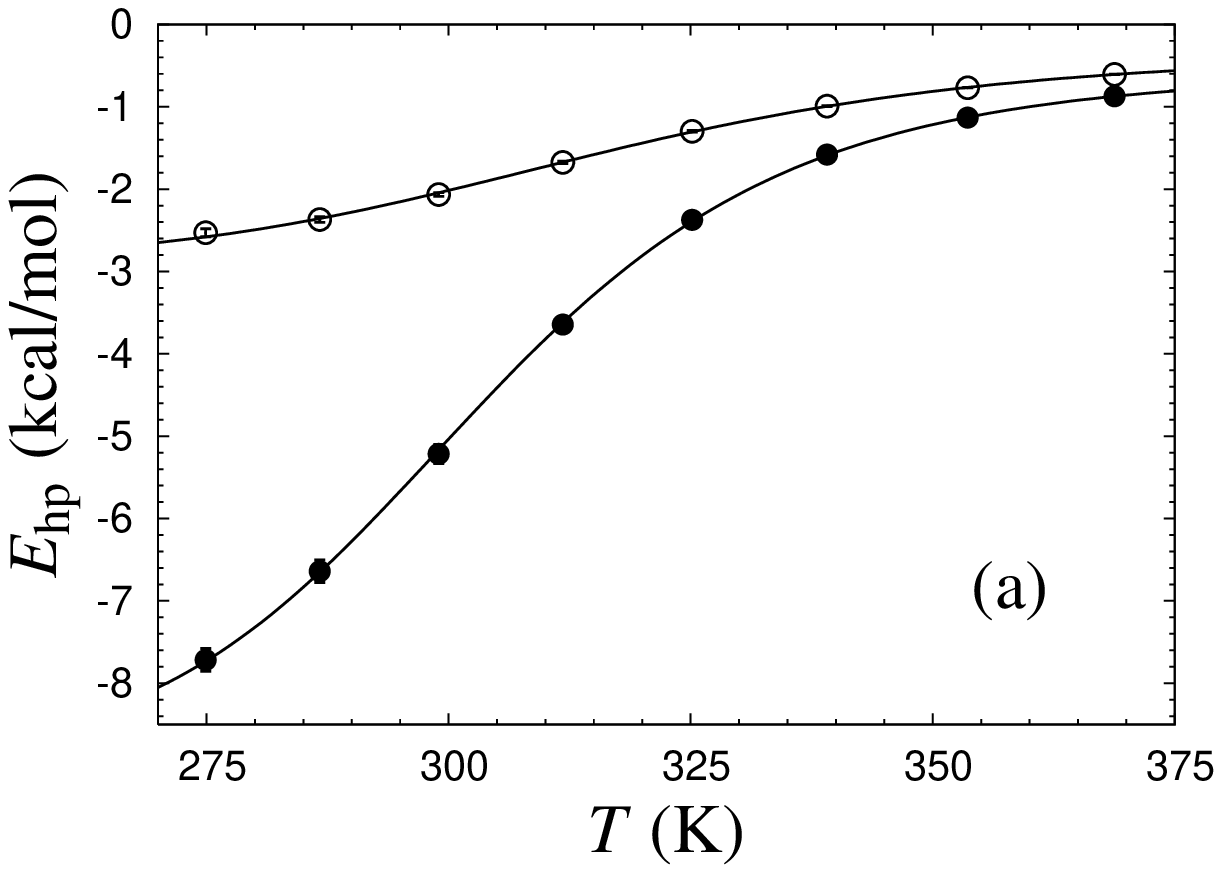,width=7cm}
\epsfig{figure=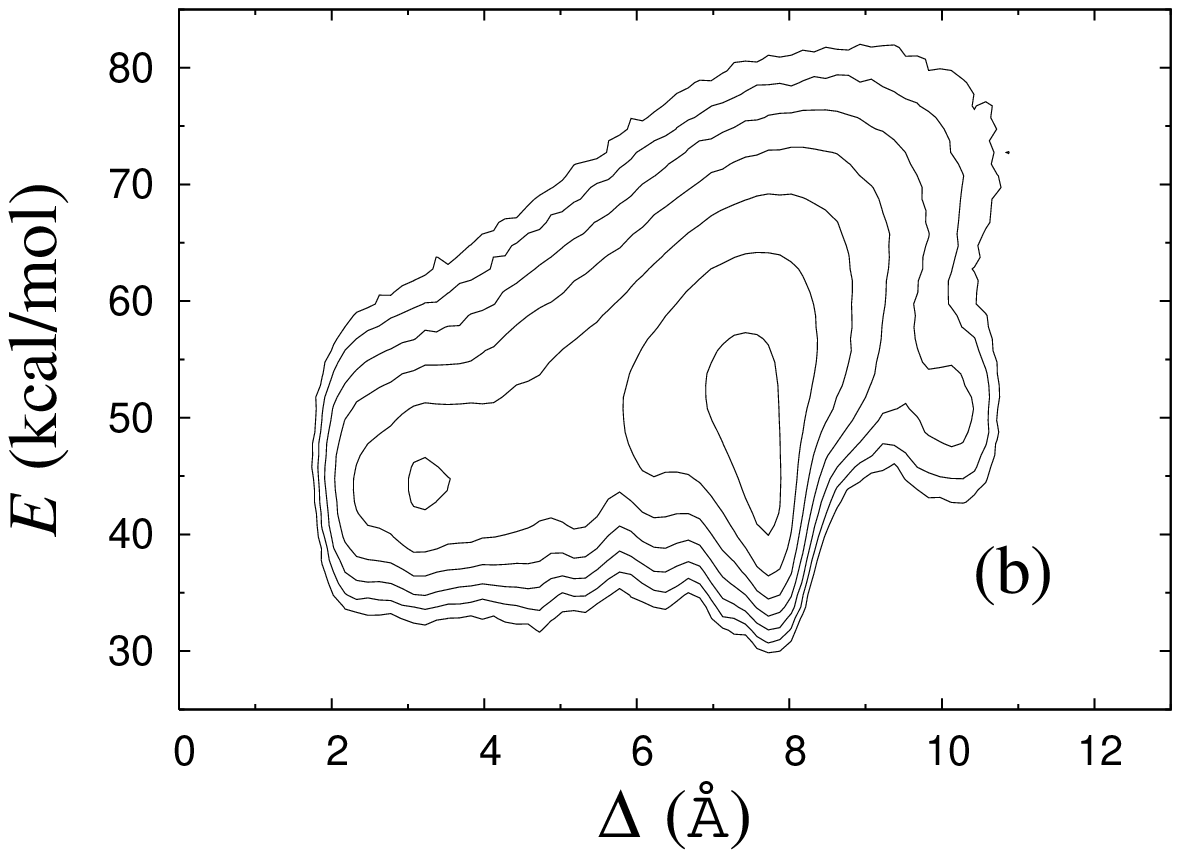,width=7cm}
\end{center}
\vspace{-3mm}
\caption{(a) The hydrophobicity energy $\Ehp$ against temperature for 
Betanova ($\circ$) and LLM ($\bullet$). 
The lines are fits to Eq.~\protect\ref{twostate}
($\Tm=314\pm1$, $\dE=8.9\pm0.1$\,kcal/mol for Betanova;
$\Tm=302\pm1$\,K, $\dE=10.9\pm0.2$\,kcal/mol for LLM).
(b) Free energy $F(\Delta,E)$ for Betanova
at 275\,K. Contour levels are as in Fig.~\protect\ref{fig:2}b.}
\label{fig:9}\end{figure}
Betanova has fewer hydrophobic residues than LLM, and we see
that $\Ehp$ is much lower in absolute value for Betanova than
for LLM. In our model, the difference in hydrophobicity is
the main reason why LLM is more stable than Betanova.
A two-state analysis of our $\Ehp$ data gives
$\Tm=314\pm1$ and $\dE=8.9\pm0.1$\,kcal/mol for Betanova,
and $\Tm=302\pm1$\,K and $\dE=10.9\pm0.2$\,kcal/mol for LLM.
These fitted two-state parameters contrast sharply
with the results of the $\Nhb$ analysis above, especially
for Betanova. In fact, for Betanova, the fitted two-state parameters
correspond to a native population of 80\,\% at the temperature 287\,K,
at which Betanova was estimated above to be only 6\,\% folded.
This discrepancy between the native populations obtained using $\Ehp$ and
$\Nhb$ data clearly show that,
in our model, these two peptides do not behave as
ideal two-state systems. It is worth noting that the quality of
the two-state fits in Fig.~9a, nevertheless, is very good,
which illustrates that deviations from the simple two-state picture
can be very hard to detect from the temperature dependence of a
single quantity~\cite{Favrin:03}.

Fig.~9b shows the free energy $F(\Delta,E)$ for Betanova at 275\,K.
Like for the $\beta$-hairpins, we use all the heavy atoms in the RMSD, but
limit the comparison to the residues 3-18. The residues 1, 2, 19 and 20
do not participate in the $\beta$-sheet structure. There is a local minimum
at $\Delta\approx3.2$\,\AA\ representing the state obtained in our model
that most resembles the NMR structure. That this state is not the most
probable state in the model is consistent with the low native population
found experimentally for this peptide. The corresponding graph
for LLM shows a much more prominent minimum representing the native
conformation.

\subsection{The character of the melting transition}

For GB1p, Betanova and LLM, we saw above that the apparent
native population depends on which quantity we study.
This dependence reflects the fact that these peptides do not
show ideal two-state behavior in the model. A quantity for which
we obtain a relatively high apparent melting temperature not
only for these three peptides but for all the peptides studied,
is the radius of gyration, $\Rg$. The $\Tm$ values obtained from our
$\Rg$ data for \Fs\ and the Trp cage are 29\,K and 9\,K higher,
respectively, than what we found above using the helix content.
For GB1m3, our $\Rg$ data gives a $\Tm$ that is 6\,K higher than
that obtained above using the hydrophobicity energy. These comparisons
show that none of the peptides studied behaves as a perfect two-state
system in our model, although the deviations from this behavior might be
relatively small for some of them, such as GB1m3.

One measure of the sharpness of the melting transition is
the height of the peak in the specific heat, $\Cv$. In Fig.~10,
we show specific heat curves for the different peptides studied.
\begin{figure}[t]
\begin{center}
\epsfig{figure=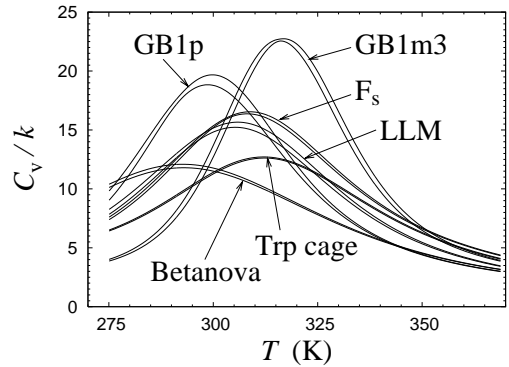,width=7cm}
\end{center}
\vspace{-3mm}
\caption{The specific heat $\Cv$ against temperature for the
different peptides, as obtained using  
histogram reweighting techniques~\protect\cite{Ferrenberg:88}. 
For each peptide, a band is shown.
The band is centered around the expected value and shows statistical
1$\sigma$ errors. $\Cv$ is defined as
$\Cv=N^{-1}d\ev{E}/dT= (NkT^2)^{-1}(\ev{E^2}-\ev{E}^2)$,
where $N$ is the number of amino acids and $\ev{O}$ denotes a 
Boltzmann average of variable $O$.}
\label{fig:10}\end{figure}
The results for GB1m2 are again very similar to those for GB1m3 and
therefore omitted.  The specific heat exhibits a clear peak
for all the peptides studied, but the height of the peak varies.
The peak is highest for GB1m3, indicating that the melting transition
is most two-state-like for this peptide. A comparison of the
energy distributions of the different peptides (not shown) supports
this conclusion. For GB1m3, we find that the energy distribution
has a bimodal shape, although not very pronounced. The other peptides
all have wide but single-peaked distributions. The distribution
is particularly wide, virtually flat, for GB1p, which has the next
highest peak in $\Cv$.

For the peptide with the sharpest transition, GB1m3, we find that the
specific heat maximum, 316\,K, is located near the temperature at which
its folded population is 50\,\%. The other peptides are less than 50\,\%
folded at their specific heat maxima, especially Betanova. Betanova
was estimated above to be 6\% folded at 287\,K in the model, and has its
specific heat maximum at a temperature higher than that, 293\,K.

\section{Conclusion}

We have developed an atomic model with a simplified phenomenological
potential for folding studies of polypeptide chains, which was tested on
a set of peptides with about 20 amino acids each, namely the Trp cage,
\Fs, GB1p, GB1m2, GB1m3, Betanova and LLM. First of all, our study
shows that the model folds these different
sequences to structures similar to their experimental
structures, for one and the same choice of model parameters.
In addition, we investigated the stability and melting behavior
of the peptides. The following list is a brief summary of these calculations,
focusing on the observables expected to be correlated with the corresponding
experimental probes.
\begin{itemize}
\item  The helix content of the Trp cage shows a temperature
dependence that is in good agreement with experimental
data based on CD and NMR (see Fig.~3).
\item A two-state analysis of the helix content of \Fs\ gives $\Tm$ and $\dE$
values that are in good agreement with CD data, while the $\Tm$ value
is somewhat low compared to its IR-based value.
\item  Estimates of folded populations based on native hydrogen bond data
for the $\beta$-sheet peptides GB1p, GB1m2, GB1m3, Betanova and LLM are
in good agreement with CD- and NMR-based experimental results, as is
summarized in Table~2. Recall that the energy scale was set
using the $\alpha$-helical Trp cage.
\item  Experimentally, GB1p has been studied using Trp fluorescence as well,
which gave a folded population higher than that in Table~2.
Our results based on hydrophobicity energy data are in good
agreement with those from the Trp fluorescence study.
\end{itemize}
The model fails to reproduce the difference in folded population
between the two stable mutants of GB1p (see Table~2),
which in part may be due to the fact that Coulomb interactions between
side-chain charges are ignored; GB1m3 contains some
charged residues that are missing in GB1m2. The overall quantitative
agreement with experimental data is, nevertheless, excellent.
This agreement indicates that factors such as Coulomb interactions
between charged residues play a quite limited role in the folding
thermodynamics of these peptides, compared to hydrogen bonding and
hydrophobic attraction, which are the main driving forces of the
model.

\begin{table}[t]
\begin{center}
\begin{tabular}{lll}
          & Exp. & Model \\
\hline
GB1p& $\sim$\,30\,\% (298\,K) & $27\pm2$\,\% (299\,K)\\
GB1m2& $74\pm5$\,\% (298\,K) & $84\pm1$\,\% (299\,K)\\
GB1m3& $86\pm3$\,\% (298\,K) & $82\pm1$\,\% (299\,K)\\
Betanova& 9\,\% (283\,K) & $6\pm1$\,\% (287\,K)\\
LLM & 36\% (283\,K)& $38\pm2$\,\% (287\,K) 
\end{tabular}
\caption{Folded populations of the different $\beta$-sheet peptides in the 
model, along with experimental results. The experimental data on GB1p, GB1m2
and GB1m3 are from Fesinmeyer et al.~\cite{Fesinmeyer:04}, whereas those on 
Betanova and LLM are from L\'opez de la Paz et al.~\cite{Lopez:01}.} 
\label{tab:2}
\end{center}
\end{table}

The temperature dependence of the model is, to us, surprisingly good, for
two reasons. First, the temperature dependence was not considered
at all when calibrating the model, except in the determination of
the energy scale. A considerable amount of fine-tuning was required
in order to obtain proper folded structures, but no further fine-tuning
was performed once that goal had been achieved. Second, our calculations
do not involve any reparametrization of the energy function. In other words,
the parameters of the energy function are temperature independent, which
is a simplifying assumption rather than a controlled approximation. On the
other hand, it should be noted that the melting transition is not
triggered by a sudden change in, for example, the strength of the
hydrophobic attraction.

In the development of this model, we have taken a purely
phenomenological approach. The model will be further developed by
studying new amino acid sequences, which will impose new
conditions on the interaction potential. As before, the challenge
will be to do this in a backwards compatible manner; the model must
not lose its ability to fold previously studied sequences.
As to limitations of the current version of the model, we
know that it is unable to properly fold the so-called trpzip
$\beta$-hairpins~\cite{Cochran:01}, which make $\beta$-hairpins
in the model but with the wrong topology. We also expect that
refinement of the model will be needed as the chains get larger.
For example, as mentioned earlier, it is likely that our pair-wise
additive hydrophobicity potential will have to be supplemented
with multibody terms for large chains. Finding out how to change
the model in order to make it more general without losing
computational efficiency will not be an easy task, but the
results obtained so far makes it tempting to try.

\subsection*{Acknowledgments}

We thank Garry Gippert for valuable discussions and
Luis Serrano and Manuela L\'opez de la Paz for providing NMR
structures for LLM and Betanova.
This work was in part supported by the Swedish Research Council
and the Knut and Alice Wallenberg Foundation through the Swegene consortium.

\end{document}